\documentclass[prl,showpacs,altaffilletter,twocolumn,
superscriptaddress,floatfix,nofootinbib,preprintnumbers]{revtex4-1} 

\pdfoutput=1
\usepackage{amsfonts,amsmath,graphicx,bm} 
\usepackage{dcolumn}
\usepackage[pdftex,
       colorlinks=true,
       urlcolor=blue,       
       linkcolor=red,       
       pdfpagemode=None,
       bookmarksopen=true]{hyperref}
\usepackage{ifpdf}
\usepackage{multirow}
\usepackage[colorlinks=true]{hyperref}
\usepackage{url}

\newcommand{\glasgow}{SUPA, School of Physics and Astronomy,
  University of Glasgow, Glasgow, G12 8QQ, UK}

\newcommand{\INFN}{INFN,  Sezione  di  Roma  Tor  Vergata,  Via  della  Ricerca  Scientifica  1,  00133  Roma  RM,  Italy}

\newcommand{\rjpsi}{0.2582(38)}
\newcommand{\gammamu}{1.73(12)\times 10^{13} \,s^{-1}}
\newcommand{\gammamugev}{11.36(81)\times 10^{-12} \,\mathrm{GeV}}
\newcommand{\gammatau}{4.45(30)\times 10^{12} \,s^{-1}}
\newcommand{\gammataugev}{2.93(19)\times 10^{-12} \,\mathrm{GeV}}

\def\today{\number\day\space\ifcase\month\or
January\or February\or March\or April\or May\or June\or
July\or August\or September\or October\or November\or December\fi
\space\number\year}
\newcount\mins \newcount\hours
\def\now{\hours=\time \mins=\time
	\divide\hours by60 \multiply\hours by60 \advance\mins by-\hours
	\divide\hours by60 
	\number\hours:\ifnum\mins<10 0\fi\number\mins }

\preprint{}

\begin{document}
\title{$R(J/\psi)$ and $B_c^- \rightarrow J/\psi \ell^-\overline{\nu}_\ell$ Lepton Flavor Universality Violating Observables from Lattice QCD}
\author{Judd \surname{Harrison}}
\email[]{judd.harrison@glasgow.ac.uk}
\affiliation{\glasgow}
\author{Christine~T.~H.~\surname{Davies}} 
\email[]{christine.davies@glasgow.ac.uk}
\affiliation{\glasgow}
\author{Andrew \surname{Lytle}} 
\affiliation{\INFN}
\collaboration{HPQCD Collaboration}
\email[]{http://www.physics.gla.ac.uk/HPQCD}
\begin{abstract}
We use our lattice QCD computation of the $B_c\rightarrow J/\psi$ form factors to determine the differential decay rate for the semitauonic decay channel and construct the ratio of branching fractions $R(J/\psi) = \mathcal{B}(B_c^- \rightarrow J/\psi \tau^-\overline{\nu}_\tau)/\mathcal{B}(B_c^- \rightarrow J/\psi \mu^-\overline{\nu}_\mu)$. 
We find $R(J/\psi) = \rjpsi$ and give an error budget. We also extend the relevant angular observables, which were recently suggested for the study of lepton flavor universality violating effects in $B\rightarrow D^*\ell\nu$, to $B_c \rightarrow J/\psi\ell\nu$ and make predictions for their values under different new physics scenarios.
\end{abstract}
\maketitle
\section{Introduction}
\label{sec:intro} 

$B$-meson exclusive semileptonic decay processes are powerful tests of the Standard model (SM) 
in cases where both experimental and theoerical uncertainties can be brought under control. 
They allow the determination of elements of the Cabibbo-Kobayashi-Maskawa matrix and tests 
of its unitarity but are also more detailed probes for new physics scenarios. 
An exciting possibility is that of non-universality of the couplings of charged 
leptons in electroweak interactions, hinted at in measurements of the 
ratio, $\mathcal{R}$, 
of the branching fraction to $\tau$ in the final state to that of $\mu/e$
for $B \rightarrow D^{(*)}$~\cite{PhysRevLett.109.101802,PhysRevD.88.072012,Huschle:2015rga,Sato:2016svk,Hirose:2016wfn,Hirose:2017dxl,Aaij:2015yra,Aaij:2017uff,Aaij:2017deq,Abdesselam:2019dgh},
\begin{equation}\label{eq:R}
\mathcal{R}(D^{(*)}) = \frac{\mathcal{B}\left(B \rightarrow D^{(*)}\tau\bar{\nu}_\tau\right)}{\mathcal{B}\left(B \rightarrow D^{(*)}\mu\bar{\nu}_\mu\right)}.
\end{equation}
The combination of experimental results 
for $\mathcal{R}(D)$ and $\mathcal{R}(D^*)$ is in tension with the SM 
at 3.1$\sigma$~\cite{HFLAV, Amhis:2019ckw}. 
This includes a recent result from Belle~\cite{Abdesselam:2019dgh} using semileptonic tagging
which by itself agrees with the SM within 1.6$\sigma$. Most of the pull comes from 
$\mathcal{R}(D^*)$ for which the SM results being used~\cite{Bernlochner:2017jka,Bigi:2017jbd,Jaiswal:2017rve} rely on 
arguments from sum rules and heavy quark effective theory because lattice QCD 
results, which provide the best {\it ab initio} determination of the form factors, 
are as yet only available for the $A_1$ form factor at zero 
recoil~\cite{Bailey:2014tva,Harrison:2017fmw}. 
Extending the lattice QCD calculation to cover more of the $q^2$ range is underway~\cite{Vaquero:2019ary, Lytle:2020tbe}. 

This motivates a study of $\mathcal{R}$ for other $b \rightarrow c$ semileptonic 
decay modes and LHCb recently gave first results 
for $\mathcal{R}(J/\psi)$ from $B_c \rightarrow J/\psi$ decay~\cite{PhysRevLett.120.121801}. 
They found $R(J/\psi)=0.71\pm0.17_\text{stat}\pm0.18_\text{sys}$ and compared to SM results 
from a variety of model calculations. Since it is difficult to quantify uncertainties 
from individual models, the spread of results between models must be taken 
as a measure of this. LHCb quote a range of SM values from 0.25 to 0.28 
from~\cite{Anisimov:1998uk,Kiselev:2002vz,Ivanov:2006ni,Hernandez:2006gt}, giving 
a 10\% spread. Since this is the same level as the discrepancy seen in $\mathcal{R}(D^*)$ 
currently, it is unlikely that this accuracy would be good enough to 
distinguish experiment (given
improved uncertainties there) and the SM. More recent results see a similar spread~\cite{Leljak:2019eyw,Issadykov:2018myx,Wang:2018duy}. In addition the dominant systematic 
error for the experiment comes from the uncertainty in input form factors 
for $B_c \rightarrow J/\psi$ from these models.  

In~\cite{longpaper} we have given results for the $B_c \rightarrow J/\psi$ form factors 
from lattice QCD for the first time. We work on gluon field configurations that include 
$u$, $d$, $s$ and $c$ quarks in the sea and use a discretisation of the Dirac equation which 
was developed by the HPQCD collaboration to have particularly small discretisation 
errors~\cite{PhysRevD.75.054502}. This makes it suitable for handling heavy $b$ and $c$ 
quarks where discretisation errors are relatively large. 
It also enables us to use a relativistic 
formalism in which the quark currents that couple to the $W$ boson can be normalised in 
a fully nonperturbative way. We use HPQCD's `heavy-HISQ' technique which has been successful 
in determining the decay constants of heavy-light 
mesons~\cite{Davies:2010ip,McNeile:2011ng,McNeile:2012qf,Bazavov:2017lyh} 
and is now being applied to $B$ 
semileptonic form factors~\cite{EuanBsDsstar,EuanBsDs} where it enables the full $q^2$ range 
of the decay to be covered. Further details can be found in~\cite{longpaper} where we give 
the form factors and derive the total rate and branching fraction 
for the decay with $\mu\overline{\nu}_{\mu}$ in the final state. Here we focus on 
the $\tau \overline{\nu}_{\tau}$ mode and determine $\mathcal{R}(J/\psi)$ in the SM with 
quantified uncertainties. We also give values for asymmetries which show sensitivity 
to new physics. 

\begin{figure}
\includegraphics[scale=0.3]{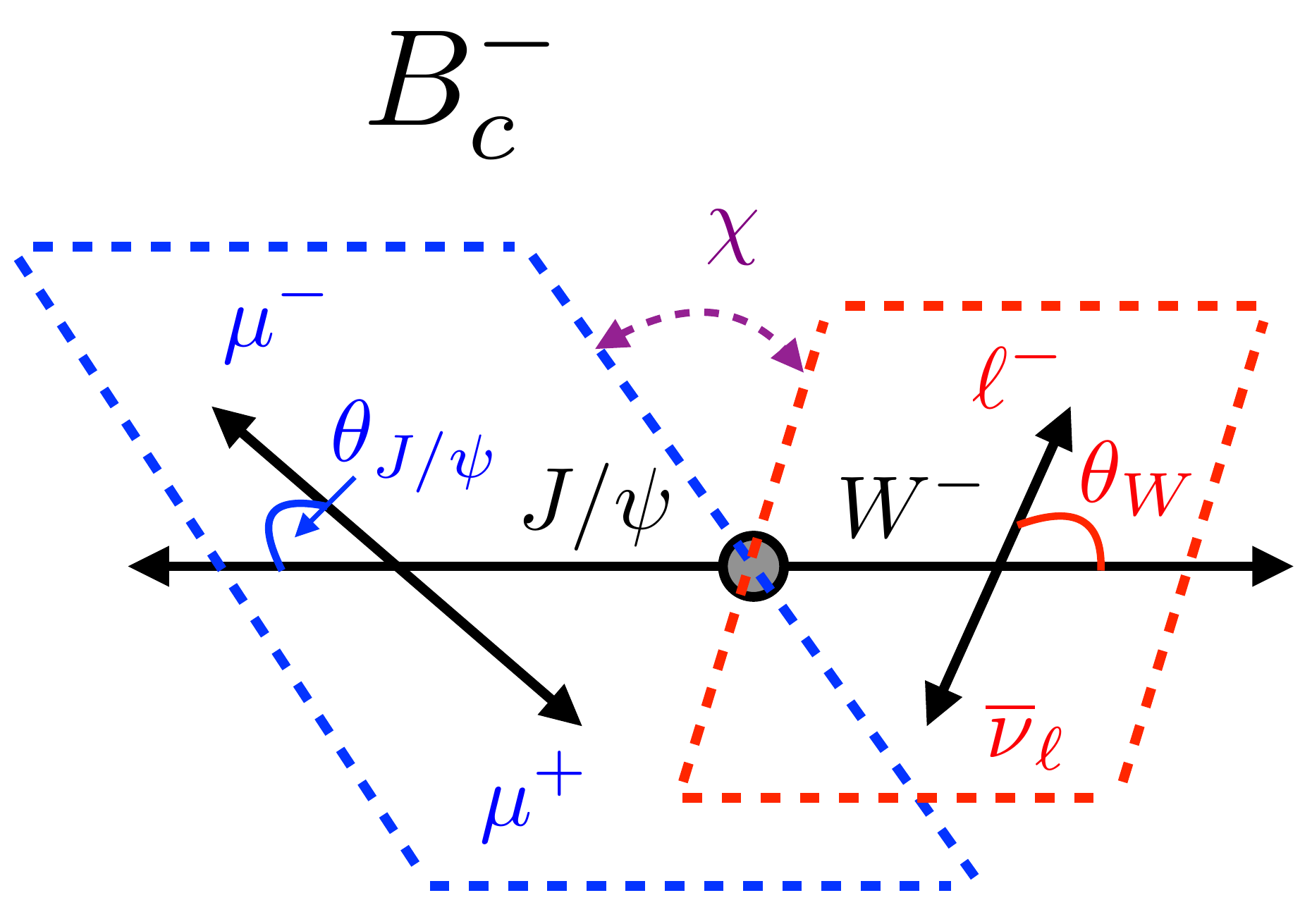}
\caption{\label{angles}Conventions for the angular variables entering the differential decay rate.}
\end{figure}

\section{Theoretical Background}
\label{sec:theory}

\begin{table}
\centering
\caption{The helicity amplitude combinations and coefficients for them that 
appear in the differential rate, Eq.~(\ref{eq:diffrate}).\label{tab:diffterms}}
\begin{tabular}{ c c | c  }
\hline
$i$ & $\mathcal{H}_i$ & $k_i(\theta_W,\theta_{J/\psi},\chi)$\\
\hline
1 & $|H_+(q^2)|^2$ & $\frac{1}{2}(1-\cos(\theta_W))^2(1+\cos^2(\theta_{J/\psi}))$\\
2 & $|H_-(q^2)|^2$ & $\frac{1}{2}(1+\cos(\theta_W))^2(1+\cos^2(\theta_{J/\psi}))$\\
3 & $|H_0|^2$& $2\sin^2(\theta_W)\sin^2(\theta_{J/\psi})$\\
4 & $\mathrm{Re}(H_+H_0^*)$&$\sin(\theta_W)\sin(2\theta_{J/\psi})\cos(\chi)(1-\cos(\theta_W))$\\
5 & $\mathrm{Re}(H_-H_0^*)$&$-\sin(\theta_W)\sin(2\theta_{J/\psi})\cos(\chi)(1+\cos(\theta_W))$\\
6 & $\mathrm{Re}(H_+H_-^*)$& $\sin^2(\theta_W)\sin^2(\theta_{J/\psi})\cos(2\chi)$\\
7 & $\frac{m_\ell^2}{q^2}|H_+(q^2)|^2$ & $\frac{1}{2}(1-\cos^2(\theta_W))(1+\cos^2(\theta_{J/\psi}))$\\
8 & $\frac{m_\ell^2}{q^2}|H_-(q^2)|^2$ & $\frac{1}{2}(1-\cos^2(\theta_W))(1+\cos^2(\theta_{J/\psi}))$\\
9 & $\frac{m_\ell^2}{q^2}|H_0|^2$& $2\cos^2(\theta_W)\sin^2(\theta_{J/\psi})$\\
10 & $\frac{m_\ell^2}{q^2}|H_t(q^2)|^2$ & $2\sin^2(\theta_{J/\psi})$\\
11 & $\frac{m_\ell^2}{q^2}\mathrm{Re}(H_+H_0^*)$&$\sin(\theta_W)\sin(2\theta_{J/\psi})\cos(\chi)\cos(\theta_W)$\\
12 & $\frac{m_\ell^2}{q^2}\mathrm{Re}(H_-H_0^*)$&$\sin(\theta_W)\sin(2\theta_{J/\psi})\cos(\chi)\cos(\theta_W)$\\
13 & $\frac{m_\ell^2}{q^2}\mathrm{Re}(H_+H_-^*)$& $-\sin^2(\theta_W)\sin^2(\theta_{J/\psi})\cos(2\chi)$\\
14 & $\frac{m_\ell^2}{q^2}\mathrm{Re}(H_tH_0^*)$& $-4\sin^2(\theta_{J/\psi})\cos(\theta_W)$\\
15 & $\frac{m_\ell^2}{q^2}\mathrm{Re}(H_+H_t^*)$& $-\sin(\theta_W)\sin(2\theta_{J/\psi})\cos(\chi)$\\
16 & $\frac{m_\ell^2}{q^2}\mathrm{Re}(H_-H_t^*)$& $-\sin(\theta_W)\sin(2\theta_{J/\psi})\cos(\chi)$
\end{tabular}
\end{table}

We give below the full differential rate~\cite{Cohen:2018vhw} for 
$B_c^- \rightarrow J/\psi \ell^- \overline{\nu}$, where $\ell$ is a 
lepton with mass $m_\ell$, with respect to 
squared 4-momentum transfer, $q^2$, and angles defined in Figure~\ref{angles}. 
We assume that the $J/\psi$ is identified through its (purely electromagnetic) 
decay to $\mu^+\mu^-$, 
defining the angle $\theta_{J/\psi}$, and we sum over the $\mu^+\mu^-$ helicities. 
\begin{align}
\label{eq:diffrate}
\frac{d^4\Gamma(B_c^-\rightarrow J/\psi(\rightarrow\mu^+\mu^-)\ell^-\overline{\nu})}{d\cos(\theta_{J/\psi})d\cos(\theta_{W})d\chi dq^2}&=\nonumber\\
 \mathcal{B}(J/\psi\rightarrow\mu^+\mu^-)\mathcal{N}\sum_ik_i&(\theta_W,\theta_{J/\psi},\chi)\mathcal{H}_i(q^2) 
\end{align}
where
\begin{equation}
\mathcal{N}=\frac{G_F^2}{(4\pi)^4}|\eta_{EW}|^2|V_{cb}|^2\frac{ 3(q^2-m_\ell^2)^2|\vec{p^\prime}| }{8M_{B_c}^2 q^2}
\end{equation}
Here $|\vec{p'}|$ is the magnitude of 
the $J/\psi$ spatial momentum in the $B_c$ rest frame
and $\eta_\mathrm{EW}$ is the same structure-independent electroweak 
correction as in~\cite{longpaper}, 1.0062(16)~\cite{Sirlin:1981ie}. 
The $k_i$ and $\mathcal{H}_i$ are given in Table~\ref{tab:diffterms}. 
We include in the expressions 
terms with factors of $m_\ell^2/q^2$ that were dropped in~\cite{longpaper}. 
These are significant for the case when $\ell=\tau$ and include the 
helicity amplitude $H_t$ which does not otherwise appear. 
Integrating over angles, the differential rate in $q^2$ is then given by
\begin{align}
&\frac{d\Gamma}{dq^2} = \mathcal{N}\times\frac{64\pi}{9}\Big[\left({H_-}^2+{H_0}^2+{H_+}^2\right) \nonumber\\
+&\frac{{m_\ell^2}}{2q^2}{ \left({H_-}^2+{H_0}^2+{H_+}^2+3 {H_t}^2\right)}\Big],
\end{align}

The helicity amplitudes are defined in terms of standard Lorentz-invariant form 
factors~\cite{RevModPhys.67.893} as
\begin{align}
H_\pm(q^2) =& (M_{B_c}+M_{J/\psi})A_1(q^2) \mp \frac{2M_{B_c}|\vec{p'}|}{M_{B_c}+M_{J/\psi}}V(q^2),\nonumber\\
H_0(q^2) =& \frac{1}{2M_{J/\psi} \sqrt{q^2}} \Big(-4\frac{M_{B_c}^2{|\vec{p'}|}^2}{M_{B_c}+M_{J/\psi}}A_2(q^2)\nonumber\\
&  +  (M_{B_c}+M_{J/\psi})(M_{B_c}^2 - M_{J/\psi}^2 - q^2)A_1(q^2) \Big),\nonumber\\
H_t(q^2) =& \frac{2M_{B_c}|\vec{p'}|}{\sqrt{q^2}}A_0(q^2).\label{helicityamplitudes}
\end{align}

\begin{figure}
\centering
\includegraphics[scale=0.2]{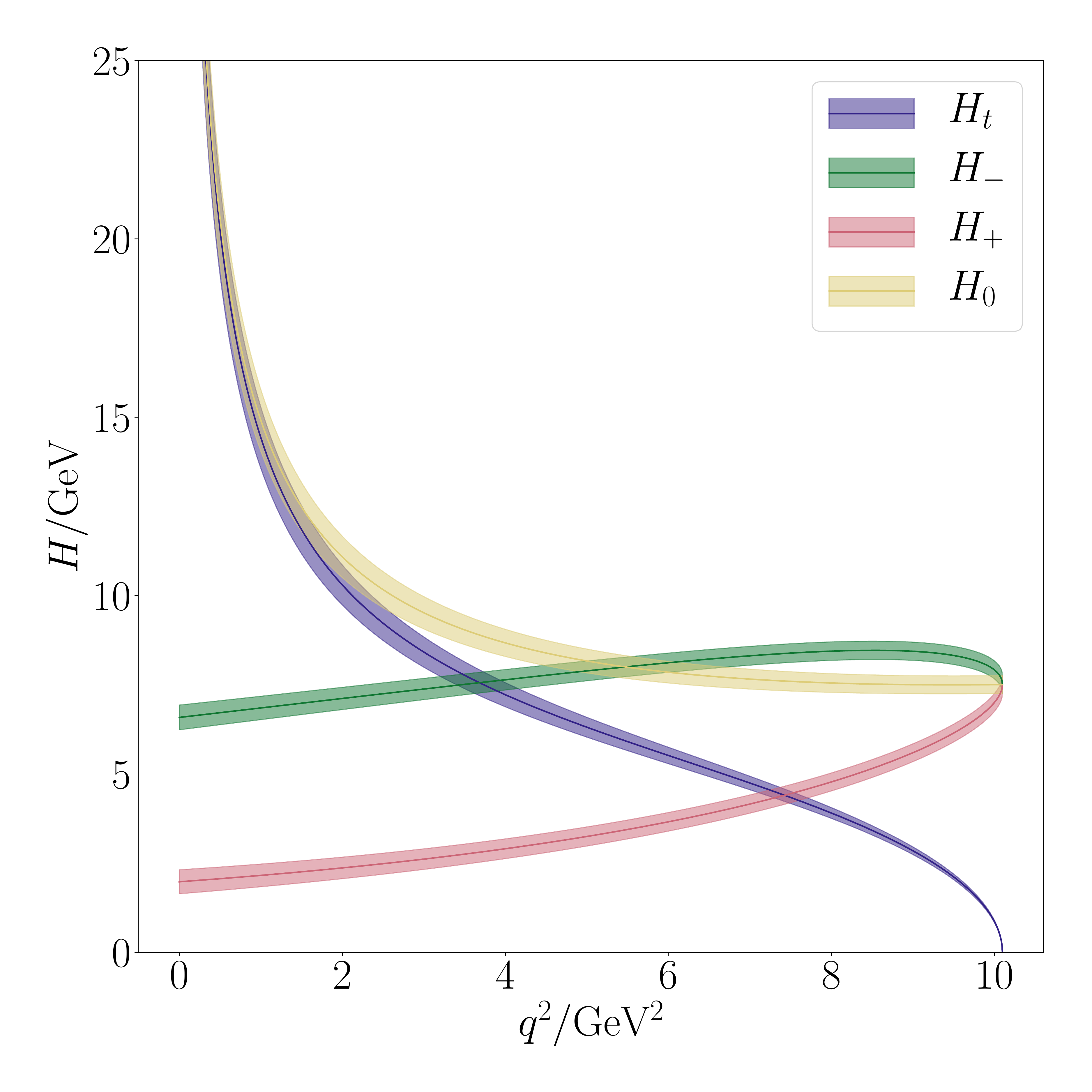}
\caption{\label{helicityamplitudesplot} Helicity amplitudes (Eq.~(\ref{helicityamplitudes})) plotted as a function of $q^2$.}
\end{figure}

\section{$\Gamma$ and $R(J/\psi)$}
\label{sec:helicityamplitudes}
The form factors were computed across the full physical $q^2$ range in~\cite{longpaper} using Lattice QCD. They are given in terms of a polynomial in $z$, with coefficients $a^F_n$, and a pole term corresponding to $B_c$ states with the quantum numbers of each current:
\begin{equation}
F(q^2) = \frac{1}{P(q^2)}\sum_{n=0}^3 a^F_n z(t_+,t_-,q^2)^n
\end{equation}
for $F=A_0,A_1,A_2,V$ and where
\begin{equation}
z(t_+,t_-,q^2) = \frac{\sqrt{t_+-q^2} - \sqrt{t_+-t_-}}{\sqrt{t_+-q^2} + \sqrt{t_+-t_-}}.
\end{equation}
$t_-$ is the maximum value of $q^2$,
$t_- = (M_{B_c}- M_{J/\psi})^2$,
$t_+$ is the pair production threshold,
$t_+ = (M_{B} + M_{D^*})^2$
and
$P(q^2) = \prod_{M_\text{pole}} z(t_+,M_\mathrm{pole},q^2)$.
The meson and subthreshold resonance masses, $M_\text{pole}$, 
that need to be used in reconstructing the form factors are given 
in~\cite{longpaper}. We assemble the 
helicity amplitudes using Eq.~(\ref{helicityamplitudes}); 
these are plotted as a function of $q^2$ in Figure~\ref{helicityamplitudesplot} 
(Figure 10 of~\cite{longpaper}).
Differential and total decay rates are then calculated.
Where an integration over $q^2$ is necessary we use a simple trapezoidal 
interpolation in order to ensure covariances are carried through correctly, 
taking sufficiently many points that the results are 
insensitive to the addition of any further points. 

\begin{figure}
\centering
\includegraphics[scale=0.2]{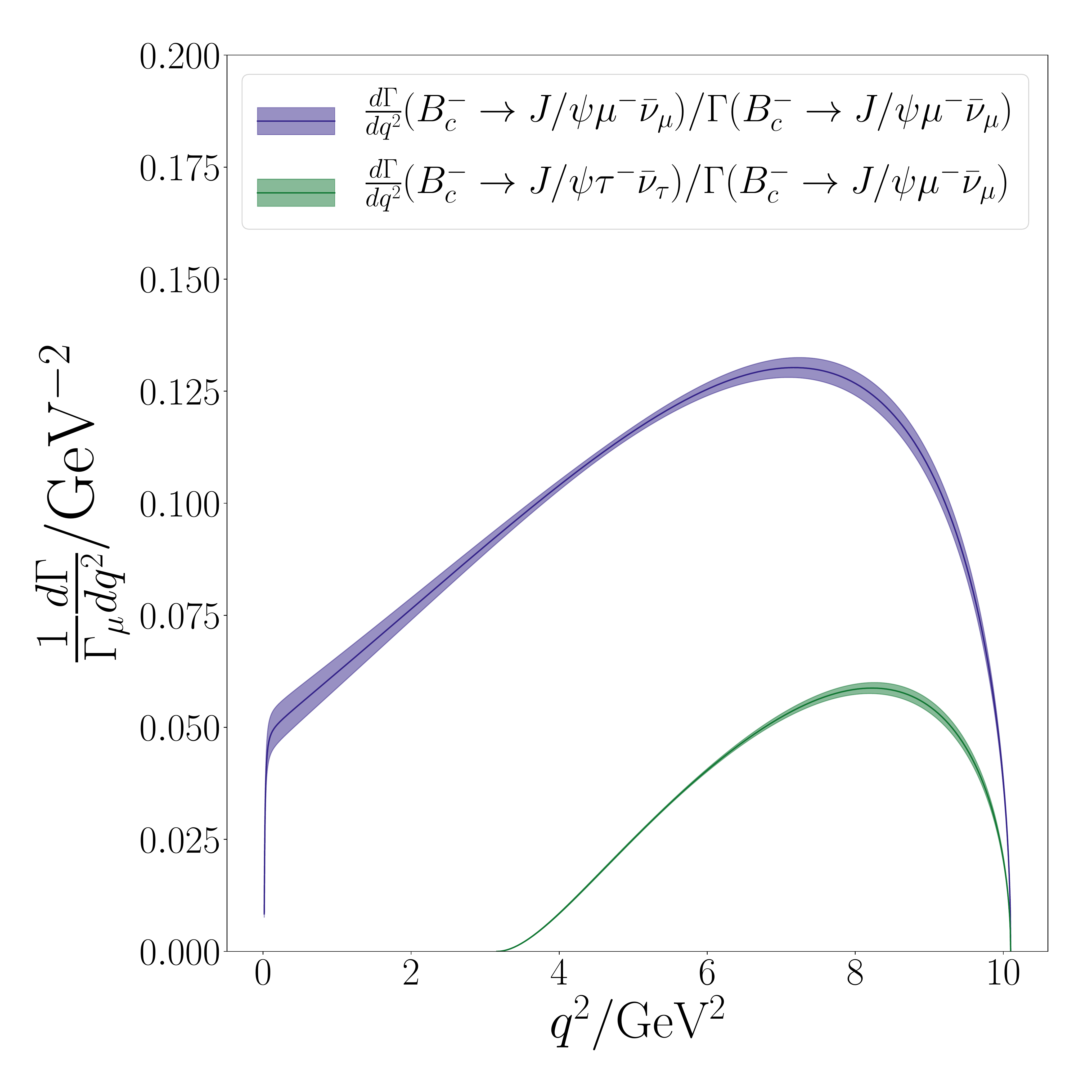}
\caption{\label{dgammadqsq} $d\Gamma/dq^2$ in the SM for the  $\ell=\mu$ and  $\ell=\tau$ cases,
normalised to the total rate for $\ell=\mu$, $\Gamma_{\mu}$.}
\end{figure}

The differential rate $d\Gamma/dq^2$ is plotted in 
Figure~\ref{dgammadqsq}, comparing SM rates for $l=\mu$ and $l=\tau$.
We also compute the total decay rates, and from 
these $R(J/\psi) = \mathcal{B}(B_c^- \rightarrow J/\psi \tau^-\overline{\nu}_\tau)/\mathcal{B}(B_c^- \rightarrow J/\psi \mu^-\overline{\nu}_\mu)$. We find
\begin{align}
\Gamma(B^-_c\rightarrow J/\psi \mu^-\overline{\nu}_\mu)/|\eta_\mathrm{EW}V_{cb}|^2 &= \gammamu\nonumber\\
&\hspace{-3.0em}=\gammamugev\nonumber\\
\Gamma(B^-_c\rightarrow J/\psi \tau^-\overline{\nu}_\tau)/|\eta_\mathrm{EW}V_{cb}|^2 &= \gammatau\nonumber\\
&\hspace{-3.0em}=\gammataugev,
\end{align}
and their ratio
\begin{equation}\label{RJPSISM}
R(J/\psi) = \rjpsi.
\end{equation}
The error budget for these results is given in Table \ref{errbudget}. 
The largest contributions for both $\Gamma(\ell=\tau)$ and $\Gamma(\ell=\mu)$ are 
the discretisation effects from the heavy quark mass, the statistical 
uncertainty in the lattice data and the quark mass mistunings effects. 
These errors and their potential improvement are discussed in~\cite{longpaper}. 
There is significant cancellation of these correlated errors in $\mathcal{R}$, 
resulting in a factor of $\approx 5$ reduction in uncertainty compared to $\Gamma$, and 
leaving the dominant error that from lattice statistics. The value for $R(J/\psi)$ is 
very close to that expected in the SM for $R(D^*)$~\cite{Amhis:2019ckw}. 
$R(J/\psi)$ is given here as the ratio of the rates to $\tau$ and $\mu$; 
we showed in~\cite{longpaper} that the decay rates to $\mu$ and $e$ differ by 
0.4\%.   

\begin{table}
\centering
\caption{Error budget for $\Gamma$ for the cases $\ell=\tau$ and $\ell=\mu$ and their ratio, $\mathcal{R}(J/\psi)$. Errors are given as a percentage of the final answer. See~\cite{longpaper} for more details.\label{errbudget}}
\begin{tabular}{ c | c c | c}
\hline
  & $\Gamma/|\eta_\mathrm{EW}V_{cb}|^2$ & &  \\\hline
Source& $\ell=\mu$ & $\ell=\tau$ & $R(J/\psi)$ \\\hline
$m_{_h}$ dependence & 2.4 & 2.2 & 0.6\\ 
continuum limit  & 3.8 & 3.6 & 0.8\\ 
sea-quark mass effects  & 3.6 & 3.4 & 0.3\\ 
lattice spacing determination & 1.4 &  1.3 & 0.1\\ 
 \hline
Statistics & 3.6 & 3.2 & 1.1 \\
 \hline
Other & 1.6 & 1.5 & 0.0  \\
 \hline
Total(\%) & 7.2 & 6.6 &  1.5\\ 
 \hline
\end{tabular}
\end{table}

\section{$R^{NP}(J/\psi)$, Angular Observables and tests of Lepton Flavor Universality}\label{angobslfu}
The effects of new physics (NP) may be considered through the inclusion of 
complex-valued NP couplings, $g_i$, $i\in{S,P,V,A,T,T5}$, in the 
effective Hamiltonian describing $b\rightarrow c\ell\nu$ decays~\cite{Bifani:2018zmi}. 
Ref.~\cite{Becirevic:2019tpx} takes
new physics to appear in the $\ell=\tau$ channel only and 
fits the $g_i$ individually against the experimental average 
values of $R(D)$ and $R(D^*)$. Angular 
observables sensitive to the different NP scenarios are then constructed. 
Here we use the values of $g$ for left-handed and right-handed vector 
couplings given in~\cite{Becirevic:2019tpx}, which we reproduce here in 
Eq.~(\ref{NPRJPSI}), and examine their impact on $R(J/\psi)$ and the 
angular observables for $B_c\rightarrow J/\psi$. 

\begin{figure*}
\includegraphics[scale=0.15]{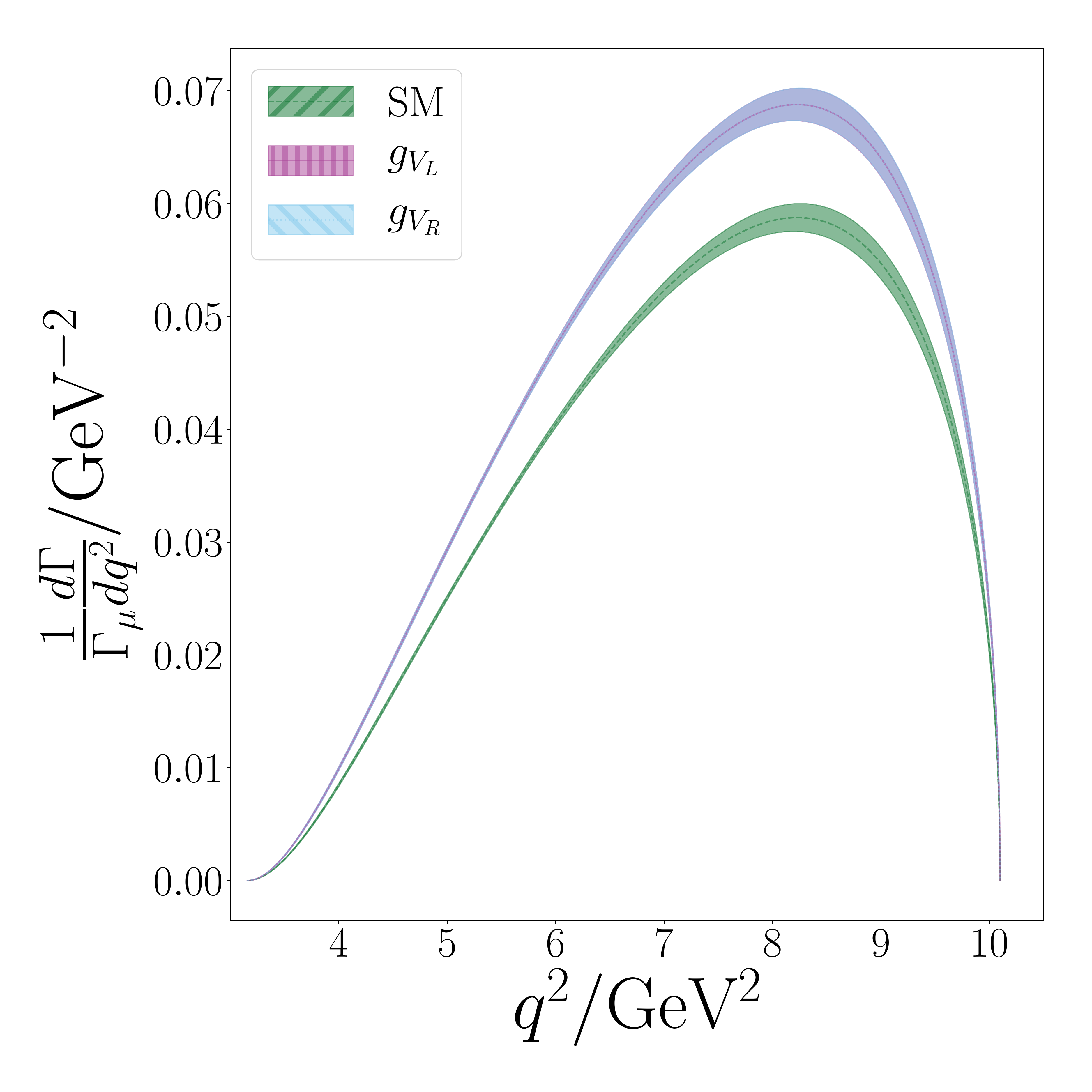}
\includegraphics[scale=0.15]{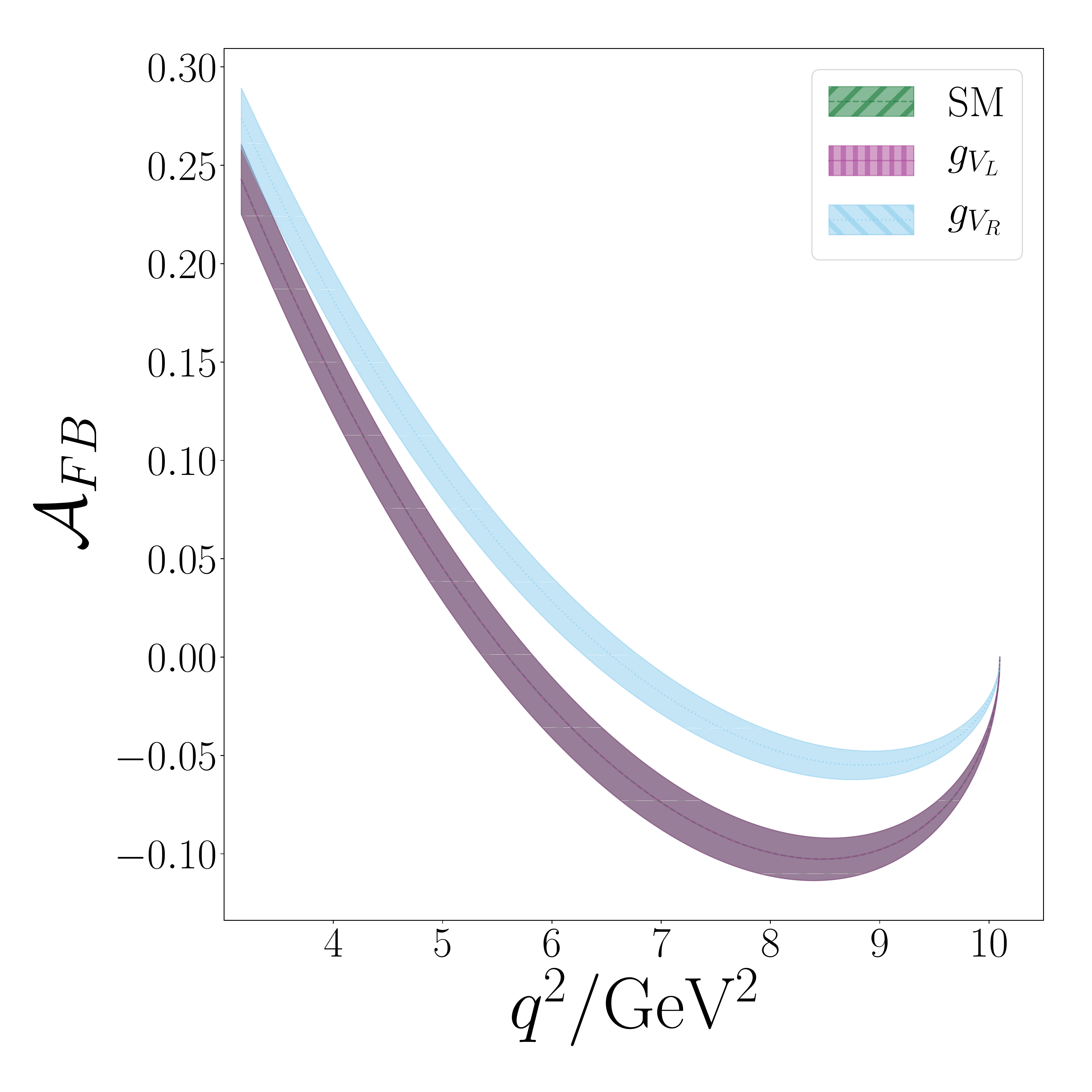}
\includegraphics[scale=0.15]{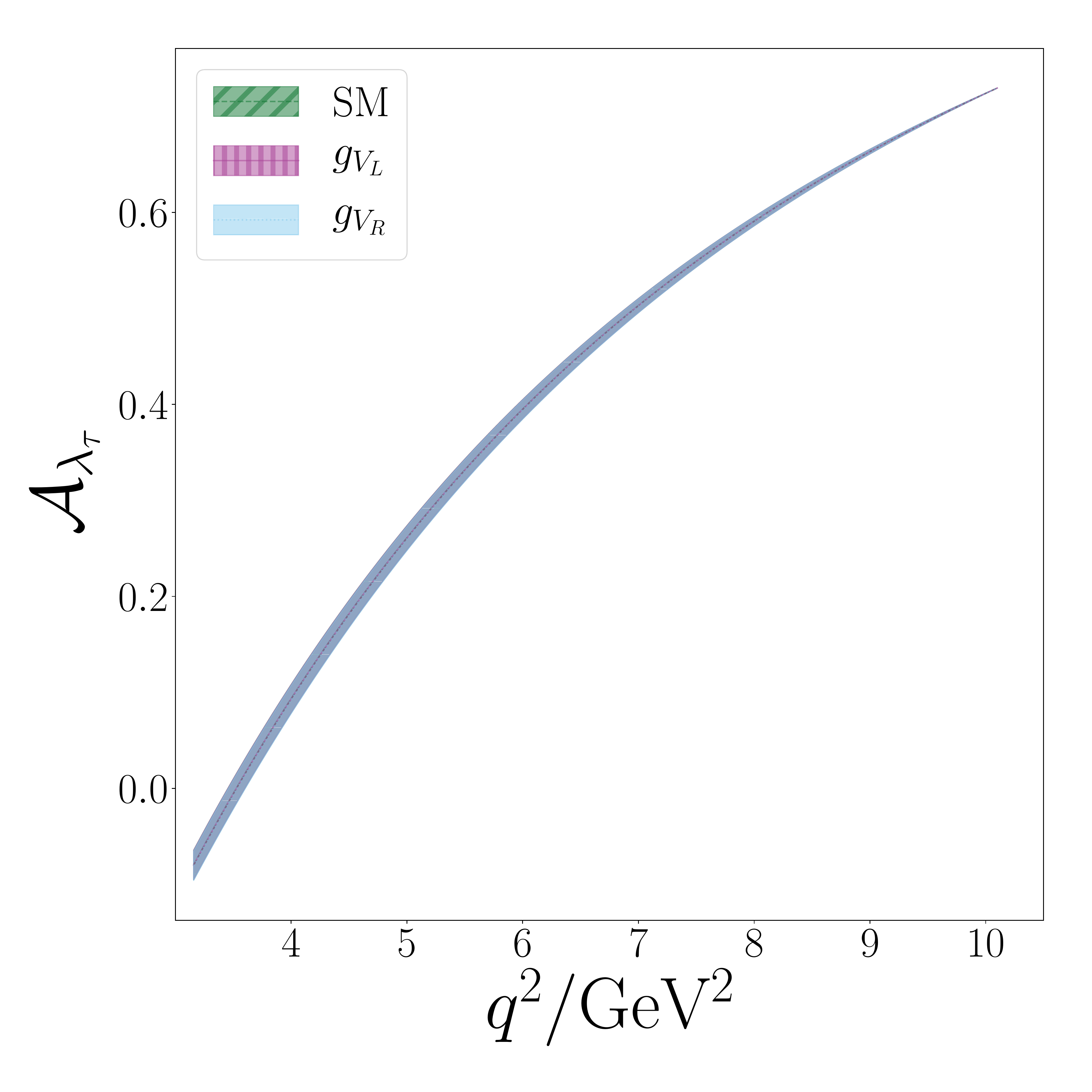}
\caption{\label{Aplot} $d\Gamma/dq^2$, $\mathcal{A}_{FB}$ and $\mathcal{A}_{\lambda_\tau}$ 
for $B_c^- \rightarrow J/\psi \tau^- \overline{\nu}_{\tau}$ in the   
SM and for the values of $g_{V_R}$ and $g_{V_L}$ given in Eq.~(\ref{NPRJPSI}) from~\cite{Becirevic:2019tpx}. $d\Gamma/dq^2$ is normalised to the total rate in the $\ell=\mu$ case, $\Gamma_{\mu}$, and the $g_{V_L}$ and $g_{V_R}$ curves overlap. For $A_{F,B}$ the SM and $g_{V_L}$ curves overlap and for $A_{\lambda_{\tau}}$ all three curves overlap. }
\end{figure*}

In Figure~\ref{Aplot} (left-hand plot) we see that the semitauonic differential rate 
increases very markedly for the best fit value of $g_{V_L}$ or $g_{V_R}$ inferred 
from $R(D^{(*)})$~\cite{Becirevic:2019tpx}. 
This results in a corresponding 10$\sigma$ increase of $R(J/\psi)$, to give 
the values below. 
\begin{align}\label{NPRJPSI}
g_{V_R} &=& -0.01 - i\,0.39;\,\,\,\,\, R^{g_{V_R}}(J/\psi)&=0.3022(44),\\
g_{V_L} &=& 0.07 - i\,0.16;\,\,\,\,\, R^{g_{V_L}}(J/\psi)&=0.3022(44), \nonumber
\end{align}

The difference between $R(J/\psi)$ and $R^{g_{V_{R/L}}}(J/\psi)$ is then close to the 
difference between the SM and current experimental average value of $R(D^*)$ that $g_{V_L}$ and $g_{V_R}$ 
were designed to reproduce. Note, however, that our values for $R^{g_{V_{R/L}}}(J/\psi)$ are still 
in tension with the experimental result.

We also compute the angular observables defined in \cite{Becirevic:2019tpx} 
which are relevant for $B_c^-\rightarrow J/\psi\ell^-\bar{\nu}_\ell$. These are 
the forward-backward asymmetry $\mathcal{A}_{FB}$ for the lepton $\ell$ 
(note that the forward direction is that of the $J/\psi$ in Figure~\ref{angles}), 
the lepton polarisation 
asymmetry $\mathcal{A}_{\lambda_\ell}$ and 
the longitudinal polarisation fraction for the $J/\psi$, $F_{L}^{J/\psi}$. Writing
\begin{align}
\frac{d^2\Gamma}{dq^2d\cos(\theta_W)} =& a_{\theta_W}(q^2)+b_{\theta_W}(q^2) \cos(\theta_W)  \nonumber\\
&+c_{\theta_W}(q^2)\cos^2(\theta_W)
\end{align}
the observables are defined as
\begin{align}
\label{eq:obs}
\mathcal{A}_{FB}(q^2) =& -\frac{b_{\theta_W}(q^2)}{d\Gamma/dq^2},\nonumber\\
\mathcal{A}_{\lambda_\ell}(q^2) =& \frac{d\Gamma^{\lambda_\ell=-1/2}/dq^2-d\Gamma^{\lambda_\ell=+1/2}/dq^2}{d\Gamma/dq^2},\nonumber\\
F_{L}^{J/\psi}(q^2) =& \frac{d\Gamma^{\lambda_{J/\psi}=0}/dq^2}{d\Gamma/dq^2}.
\end{align}
$A_{FB}$ and $A_{\lambda_{\tau}}$ 
are plotted as a function of $q^2$ in Figure~\ref{Aplot}, showing both the 
behaviour in the SM and the impact 
of the possible NP couplings $g_{V_R}$ and $g_{V_L}$. Note the very different 
shape of the SM curves for $A_{FB}$ and $A_{\lambda_{\tau}}$ in the SM 
compared to those for $\ell=e$ or $\mu$ given in~\cite{longpaper}. The helicity -1 
virtual $W$ will throw a helicity -1/2 lepton predominantly in the $W$ direction 
(i.e. backwards) but the mass of the $\tau$ changes the lepton helicity mixture. 
$g_{V_R}$ accentuates this effect by boosting the contribution of $|H_+|^2$
but without changing the $\tau$ helicity mixture.

For an observable $O_i^\ell(q^2)=N_i^\ell(q^2)/D_i^\ell(q^2)$, the integrated quantities are defined as
\begin{align}
\label{eq:int}
\langle O_i^\ell \rangle = \int_{m_\ell^2}^{q^2_\text{max}} N_i^\ell(q^2) dq^2 \Big/\int_{m_\ell^2}^{q^2_\text{max}} D_i^\ell(q^2) dq^2.
\end{align}
We give results for $\langle \mathcal{A}_{\lambda_\tau}\rangle $, 
$\langle \mathcal{A}_{FB}\rangle$ and 
$\langle F_{L}^{J/\psi}\rangle$ in Table~\ref{intangvar} for 
$B_c^- \rightarrow J/\psi \tau^- \overline{\nu}_{\tau}$ in the 
SM and for the NP couplings $g_{V_R}$ and $g_{V_L}$ from Eq.~(\ref{NPRJPSI}). 
Our SM results agree within 2$\sigma$ with those from a covariant light-front 
quark model that included information from our preliminary lattice QCD 
results~\cite{Huang:2018nnq}. 

\begin{table} 
\centering
\caption{\label{intangvar}Integrated angular observables for $B_c^-\rightarrow J/\psi \tau^- \overline{\nu}_{\tau}$ in the SM and for possible NP left-handed and right-handed vector couplings from Eq.~(\ref{NPRJPSI}).}
\begin{tabular}{ c | c c c  }
\hline
 & SM & $g_{V_R}$ & $g_{V_L}$  \\\hline
$\langle \mathcal{A}_{FB}\rangle$ & -0.058(12) & -0.0089(96) & -0.058(12) \\
$\langle \mathcal{A}_{\lambda_\tau}\rangle $ & 0.5185(75) & 0.5183(75) & 0.5185(75) \\
$\langle F_{L}^{J/\psi}\rangle $ & 0.4416(92) & 0.4423(92) & 0.4416(92) \\\hline
\end{tabular}
\end{table}

\begin{table} 
\centering
\caption{\label{lfuvtab} LFUV variables for $B_c \rightarrow J/\psi$ defined 
in Eqs~(\ref{eq:obs}),~(\ref{eq:int}) and~(\ref{eq:rat})~\cite{Becirevic:2019tpx}.
The second column gives results in the SM and then the further two columns 
give results for NP couplings, in the $\tau$ channel only, of $g_{V_R}$ and $g_{V_L}$ 
(Eq.~(\ref{NPRJPSI})). 
}
\begin{tabular}{ c | c c c  }
\hline
 & SM & $g_{V_R}$ & $g_{V_L}$ \\\hline
$R(\mathcal{A}_{FB})$ & 0.255(38) & 0.039(40) & 0.255(38) \\
$R(\mathcal{A}_{\lambda_\ell})$ & 0.5216(74) & 0.5214(74) & 0.5216(74)\\
$R(F_{L}^{J/\psi})$ & 0.887(10) & 0.889(10) & 0.887(10) \\\hline
\end{tabular}
\end{table}

We also construct the lepton flavor universality violating (LFUV) ratios 
\begin{align}
\label{eq:rat}
R(O_i) = \frac{\langle  O_i^\tau \rangle}{\frac{1}{2}(\langle O_i^\mu \rangle+ \langle O_i^e \rangle)}.
\end{align}
These are given in Table~\ref{lfuvtab} where we see that 
$R(\mathcal{A}_{FB})$ can distinguish between a NP right-handed vector coupling, 
and a left-handed one. None of the other LFUV ratios change
significantly from their SM values under the addition of either $g_{V_R}$ or $g_{V_L}$. 
This is consistent with what was seen for 
$B\rightarrow D^{*}\ell\bar{\nu}_\ell$ in~\cite{Becirevic:2019tpx}. 

\section{Conclusion}
\label{sec:conclusion}
We give the first computation in lattice QCD 
of the branching fraction ratio $R(J/\psi)$ that tests for lepton flavour universality 
in $B_c \rightarrow J/\psi$ semileptonic decay. 
Our value in the SM is $R(J/\psi) = \rjpsi$, with error budget in 
Table~\ref{errbudget}. This is in tension with the LHCb result at $1.8\sigma$ where 
$\sigma$ is the experimental uncertainty.
The $B_c \rightarrow J/\psi$ form factors that we have 
calculated~\cite{longpaper} to do this 
should enable the dominant systematic error in the experimental determination of 
$R(J/\psi)$ to be reduced, allowing progress towards an accurate test of the SM. 

We illustrate how NP might show up in $B_c \rightarrow J/\psi$ 
decay with predictions for a variety of angular observables and $\tau$ to $e/\mu$ ratios 
both in the SM and with additional NP couplings consistent with 
the current average of experimental measurements 
of $B\rightarrow D^{(*)}$ decay. We have shown that $R(J/\psi)$ is close to the value of
$R(D^*)$ both in the SM and in the presence of NP entering through either $g_{V_R}$ or $g_{V_L}$.
The resultant value of $R^{g_{V_{R/L}}}(J/\psi)$ is still in tension with the experimental value. 

\subsection*{\bf{Acknowledgements}}

We are grateful to the MILC collaboration for the use of
their configurations and code. 
We thank C. Bouchard, B. Colquhoun, J. Koponen, P. Lepage, 
E. McLean and C. McNeile for useful discussions.
Computing was done on the Cambridge service for Data 
Driven Discovery (CSD3), part of which is operated by the 
University of Cambridge Research Computing on behalf of 
the DIRAC 
HPC Facility of the Science and Technology Facilities 
Council (STFC). The DIRAC component of CSD3 was funded by 
BEIS capital funding via STFC capital grants ST/P002307/1 and 
ST/R002452/1 and STFC operations grant ST/R00689X/1. 
DiRAC is part of the national e-infrastructure.  
We are grateful to the CSD3 support staff for assistance.
Funding for this work came from the UK
Science and Technology Facilities Council grants 
ST/L000466/1 and ST/P000746/1.

\bibliographystyle{apsrev4-1}
\bibliography{BcJpsi}

\end{document}